\documentclass[conference, 9pt]{IEEEtran}
\usepackage{cite}
\usepackage{amsmath,amssymb,amsfonts}
\usepackage{algorithmic}
\usepackage{graphicx}
\usepackage{textcomp}
\usepackage{xcolor}
\def\BibTeX{{\rm B\kern-.05em{\sc i\kern-.025em b}\kern-.08em
    T\kern-.1667em\lower.7ex\hbox{E}\kern-.125emX}}
\usepackage{listings}
\definecolor{codegreen}{rgb}{0,0.6,0}
\definecolor{codegray}{rgb}{0.5,0.5,0.5}
\definecolor{codepurple}{rgb}{0.58,0,0.82}
\definecolor{backcolour}{rgb}{0.95,0.95,0.92}
\lstdefinestyle{mystyle}{
    backgroundcolor=\color{backcolour},   
    commentstyle=\color{codegreen},
    keywordstyle=\color{magenta},
    numberstyle=\tiny\color{codegray},
    stringstyle=\color{codepurple},
    basicstyle=\ttfamily\footnotesize,
    breakatwhitespace=false,         
    breaklines=true,                 
    captionpos=b,                    
    keepspaces=false,                 
    showspaces=false,                
    showstringspaces=false,
    showtabs=false,                  
    tabsize=2
}
\usepackage{caption}
\usepackage{subcaption}
\usepackage[english]{babel}
\newtheorem{theorem}{Theorem}
\newtheorem{corollary}{Corollary}[theorem]

\usepackage{hyperref}
\usepackage[utf8x]{inputenc}

\begin{document}
\bstctlcite{daniel-bochen-tan-certified:BSTcontrol}
\title{Optimal Qubit Mapping with Simultaneous Gate Absorption \\
\vspace{5pt}
\Large{ICCAD Special Session Paper}
\vspace{-12pt}}

\author{\IEEEauthorblockN{Bochen Tan}
\IEEEauthorblockA{\textit{University of California, Los Angeles}\\
bctan@cs.ucla.edu}
\and
\IEEEauthorblockN{Jason Cong}
\IEEEauthorblockA{\textit{University of California, Los Angeles} \\
cong@cs.ucla.edu}
}

\maketitle

\begin{abstract}
Before quantum error correction (QEC) is achieved, quantum computers focus on noisy intermediate-scale quantum (NISQ) applications.
Compared to the well-known quantum algorithms requiring QEC, like Shor’s or Grover’s algorithm, NISQ applications have different structures and properties to exploit in compilation.
A key step in compilation is mapping the qubits in the program to physical qubits on a given quantum computer, which has been shown to be an NP-hard problem.
In this paper, we present OLSQ-GA, an optimal qubit mapper with a key feature of simultaneous SWAP gate absorption during qubit mapping, which we show to be a very effective optimization technique for NISQ applications.
For the class of quantum approximate optimization algorithm (QAOA), an important NISQ application, OLSQ-GA reduces depth by up to 50.0$\%$ and SWAP count by 100$\%$ compared to other state-of-the-art methods, which translates to 55.9$\%$ fidelity improvement.
The solution optimality of OLSQ-GA is achieved by the exact SMT formulation.
For better scalability, we augment our approach with additional constraints in the form of initial mapping or alternating matching, which speeds up OLSQ-GA by up to 272X with no or little loss of optimality.

\end{abstract}

\section{Introduction} \label{sec:introduction}
Compared to conventional computing using semiconductors, quantum computing (QC) suffers from high gate error rates and also a temporal decay of quantum information called decoherence.
Thus, quantum error correction (QEC) and fault-tolerance are essential to run some well-known QC applications like Shor's algorithm for factoring \cite{sfcs94-shor-discrete-logarithms-factoring} or Grover's algorithm for searching \cite{stoc96-grover-search}.
Although there has been much progress in QEC research, significant improvements in gate fidelity and error mitigation are still required \cite{arxiv2101-google-suppression-phase-error}.
Another important research direction is making use of the existing \textit{noisy intermediate-scale} hardware by employing NISQ applications \cite{quantum18-preskill-nisq}.
These include the quantum approximate optimization algorithm (QAOA) \cite{algorithms19-hadfield-wang-ogorman-rieffel-venturelli-biswas-qaoa, natphys21-google-qaoa}, and chemical simulation \cite{prl18-kivlichan-mcclean-wiebe-gidney-aspuru-guzik-chan-babbush-vqe-linear-ansatz, rmp20-mcardle-endo-aspuru-guzik-benjamin-yuan-quantum-conputational-chemistry, science20-google-hartree-fock, arxiv2010-google-fermi-hubbard}.
To benchmark the NISQ QC performance, strategies like quantum volume (QV) \cite{pra19-cross-bishop-sheldon-nation-gambetta-quantum-volume} have also been developed.

To run NISQ applications, the qubits in the program have to be mapped to physical qubits on the hardware, called layout synthesis in \cite{tc20-tan-cong-optimality-layout-queko, iccad20-tan-cong-optimal-layout-synthesis}.
Many QC architectures have connectivity constraints in the form of coupling graphs between physical qubits.
The essential entangling two-qubit gates can only be applied to two adjacent qubits on the coupling graph; however, the application may require entangling gates on any pair of qubits.
Thus, the compiler has to ``move'' the required qubits together on the coupling graph, usually via SWAP gates.
Although various settings of the mapping and SWAP insertion problem have been proved NP-hard \cite{tcad08-maslov-falconer-mosca-placement, cgo18-siraichi-santos-collange-pereira-qubit-allocation, socs18-botea-kishimoto-marinescu-complexity-quantum-compilation, tc20-tan-cong-optimality-layout-queko}, given the limited QC resource in the NISQ era, we strive for optimal mapping solutions, as the high error rates and short coherence limit the circuit size and depth.

Many research works have tried to solve the mapping problem \cite{tcad08-maslov-falconer-mosca-placement, cgo18-siraichi-santos-collange-pereira-qubit-allocation, asplos19-li-ding-xie-sabre-mapping, isca19-murali-linke-martonisi-abhari-nguyen-alderete-triq-architecture-studies, date18-zulehner-paler-wille-efficient-mapping-ibmqx, dac19-wille-burgholzer-zulehner-mapping-minimal-swaph, asplos21-zhang-hayes-qiu-jin-chen-zhang-time-optimal-mapping}, but few have considered the special properties of NISQ applications \cite{arxiv2106-nannicini-bishop-gunluk-jurcevic-optimal-mapping-bip, aspdac19-zulehner-wille-su(4)-compiling}, e.g., \textit{SWAP gate absorption} by $U(4)$ gates.
Also, many works favor heuristic solutions for scalability over optimality, despite the scale of current NISQ experiments remain moderate \cite{natphys21-google-qaoa, science20-google-hartree-fock, arxiv2010-google-fermi-hubbard, qst21-ibm-qv64}.

In this paper, we present OLSQ-GA, an optimal mapper of NISQ applications that takes into consideration both commutation and SWAP absorption.
Given these new degrees of freedom, OLSQ-GA is able to outperform state-of-the-art mappers in reducing depth and SWAP count (i.e., improving fidelity) on a set of QAOA benchmarks with similar settings of a leading experimental work \cite{natphys21-google-qaoa}.
Given the NP-hardness of the mapping problem and the SMT-based exact formulation OLSQ-GA, we cannot expect to find optimal solutions very fast.
However, we can add more constraints to the formulation to reduce solution space.
We prove that, for linear architecture, optimal mapping solutions have the pattern of \textit{alternating matchings}.
By constraining the solution space with such pattern, OLSQ-GA is sped up significantly without loss of optimality.
We also introduce other constraints like setting initial mapping or alternating matchings on non-linear architectures, which speeds up the solving with little and often no loss of optimality.

The organization of the paper is as below.
In Sec.~\ref{sec:background}, we provide some background on QC, especially in NISQ setting. 
In Sec.~\ref{sec:formulation}, we present the OLSQ-GA formulation. 
In Sec.~\ref{sec:qaoa}, we evaluate OLSQ-GA against previous works on QAOA benchmarks.
In Sec.~\ref{sec:analysis}, we perform some analysis on the structure of optimal solutions and discuss our speedup strategies.
In Sec.~\ref{sec:related}, we review related works.
In Sec.~\ref{sec:conclusion}, we summarize the results and discuss future directions.

\section{Background} \label{sec:background}
\subsection{Quantum Computing}

\begin{figure}[hbt]
    \centering
    \includegraphics[scale=0.6]{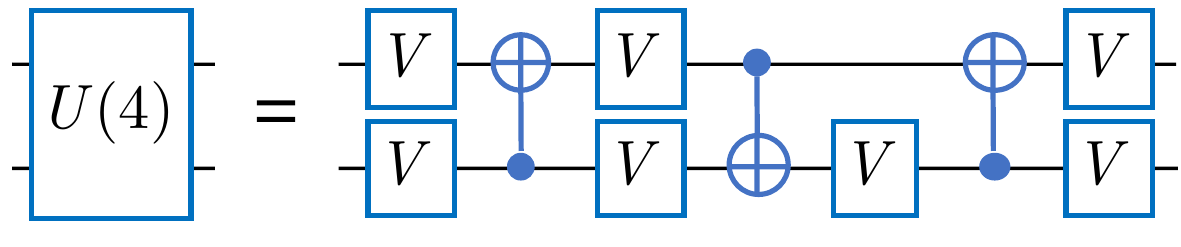}
    \caption{KAK decomposition \cite{pra04-vatan-williams-optimal-two-qubit} of $U(4)$ gate. In such QC diagrams, each wire is a qubit. Time flows from left to right. The connected $\bullet$ and $\oplus$ represent a CNOT gate. $V$s are generic single-qubit gates in Eq.~\ref{eq:U(2)}, each with its own parameters $\theta$ $\lambda$, and $\phi$.}
    \label{fig:kak}
\end{figure}

The state of a single qubit is represented as a normalized vector of length 2.
The state of $n$ qubits is then represented as a normalized vector of length $2^n$.
Quantum gates are operations that transform a state to another, so the most general $n$-qubit gate is just a $2^n$-by-$2^n$ matrix that preserves the norm, i.e., a unitary matrix.
Thus, any single-qubit gate is in the set of all 2-by-2 unitary matrix denoted as $U(2)$.
Any $U(2)$ matrix can be written as
\begin{equation} \label{eq:U(2)}
    V(\theta,\lambda,\phi)=\begin{bmatrix}
    \cos(\theta/2) &-e^{i\lambda}\sin(\theta/2) \\ e^{i\phi}\sin(\theta/2) & e^{i\lambda+i\phi}\cos(\theta/2)\\
    \end{bmatrix}.
\end{equation}
Similarly, any two-qubit gate is in the set of all 4-by-4 unitary matrix denoted as $U(4)$.
These are two common $U(4)$ gates:
\begin{equation} \label{fig:two-qubit}
\text{CNOT}=\begin{bmatrix}
    1 &0 &0 &0 \\
    0 &1 &0 &0 \\
    0 &0 &0 &1 \\
    0 &0 &1 &0 \\
    \end{bmatrix},\ 
\text{SWAP}=\begin{bmatrix}
    1 &0 &0 &0 \\
    0 &0 &1 &0 \\
    0 &1 &0 &0 \\
    0 &0 &0 &1 \\
\end{bmatrix}.
\end{equation}
Implementation of even large gates is significantly harder.
However, Ref.~\cite{book10-nielsen-chuang-quantum-computation-information} demonstrates that single-qubit and two-qubit gates are sufficient for QC, so NISQ quantum programs are usually written as a list of single-qubit and two-qubit gates.
For example, Fig.~\ref{fig:eg-program} shows a program for general chemical simulation which consists of 10 two-qubit gates on 5 qubits.
The gates in this program are fermionic simulation gates with different parameters \cite{prl20-google-fsin-gate}
\begin{equation}
    \text{fSim}(\theta,\phi) = 
    \begin{bmatrix}
    1 &0 &0 &0 \\
    0 &\cos\theta &-i\sin\theta &0 \\
    0 &-i\sin\theta &\cos\theta &0 \\
    0 &0 &0 &e^{-i\phi} \\
    \end{bmatrix}.
\end{equation}
Usually the NISQ QC hardware supports a few $U(4)$ gates but generic single-qubit gates.
E.g., IBM hardware supports the above mentioned CNOT and $V$.
To implement other $U(4)$ gates, we need to decompose them into the native $U(4)$ and some single-qubit gates.
Fig.~\ref{fig:kak} shows a commonly used KAK decomposition leveraging 3 CNOT gates, which is minimal in terms of CNOT gates \cite{pra04-vatan-williams-optimal-two-qubit}.
Ref.~\cite{micro20-gokhale-javadi-abhari-earnest-shi-chong-compilation-openpulse} provides the decomposition of a few common gates in NISQ assuming different native gates.
Since the decomposition is a purely local process, we can perform it after solving the mapping problem.

\subsection{Mapping Quantum Programs to Hardware} \label{ssec:mapping-problem}
\begin{figure}[bht]
    \centering
    \begin{subfigure}[b]{\linewidth}
    \begin{lstlisting}
g0(q0, q1);  g1(q0, q2);  g2(q0, q3);  g3(q0, q4);
g4(q1, q2);  g5(q1, q3);  g6(q1, q4);
g7(q2, q3);  g8(q2, q4);  g9(q3, q4);
    \end{lstlisting}
    \caption{A general chemical simulation on 5 qubits. The quantum program is read from left to right, and from top to bottom.}
    \label{fig:eg-program}
    \end{subfigure}
    \vfill
    \begin{subfigure}[b]{\linewidth}
    \centering
    \includegraphics[scale=0.6]{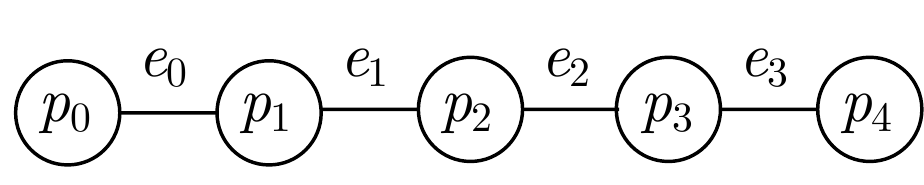}
    \caption{The coupling graph of a linear architecture to run simulation.}
    \label{fig:eg-arch}
    \end{subfigure}
    \caption{The layout/mapping problem in QC.}
    \label{fig:eg-problem}
\end{figure}

If we run the chemical simulation program shown in Fig.~\ref{fig:eg-program} on a linear QC architecture such the one in Fig.~\ref{fig:eg-arch}, the native two-qubit gate can only be applied to adjacent physical qubits on the coupling graph of the architecture.
Note that qubits in the program, $Q=\{q_i|i=0,..., 4\}$, is different from physical qubits on the architecture, $P=\{p_i|i=0,..., 4\}$.
The former is only a symbol used when writing the program, whereas the latter refers to a physical entity on the chip.
We can observe in Fig.~\ref{fig:eg-program} that $g_{0}$ is on $q_0$ and $q_1$, $g_{1}$ is on $q_0$ and $q_2$, $g_{4}$ is on $q_1$ and $q_2$.
If there is only a static mapping $\pi:Q\to P$ for the whole program, then $\pi(q_0)$, $\pi(q_1)$, and $\pi(q_2)$ should all be adjacent because $g_{0}$, $g_{1}$, and $g_{4}$ are all applied successfully.
This means that there is a triangular connection on the architecture, which contradicts with Fig.~\ref{fig:eg-arch}.
Thus, we begin with some initial mapping, and change the mapping dynamically in the execution of the program.

\begin{figure}[bt]
    \centering
    \begin{subfigure}[b]{\linewidth}
    \centering
        \includegraphics[scale=0.6]{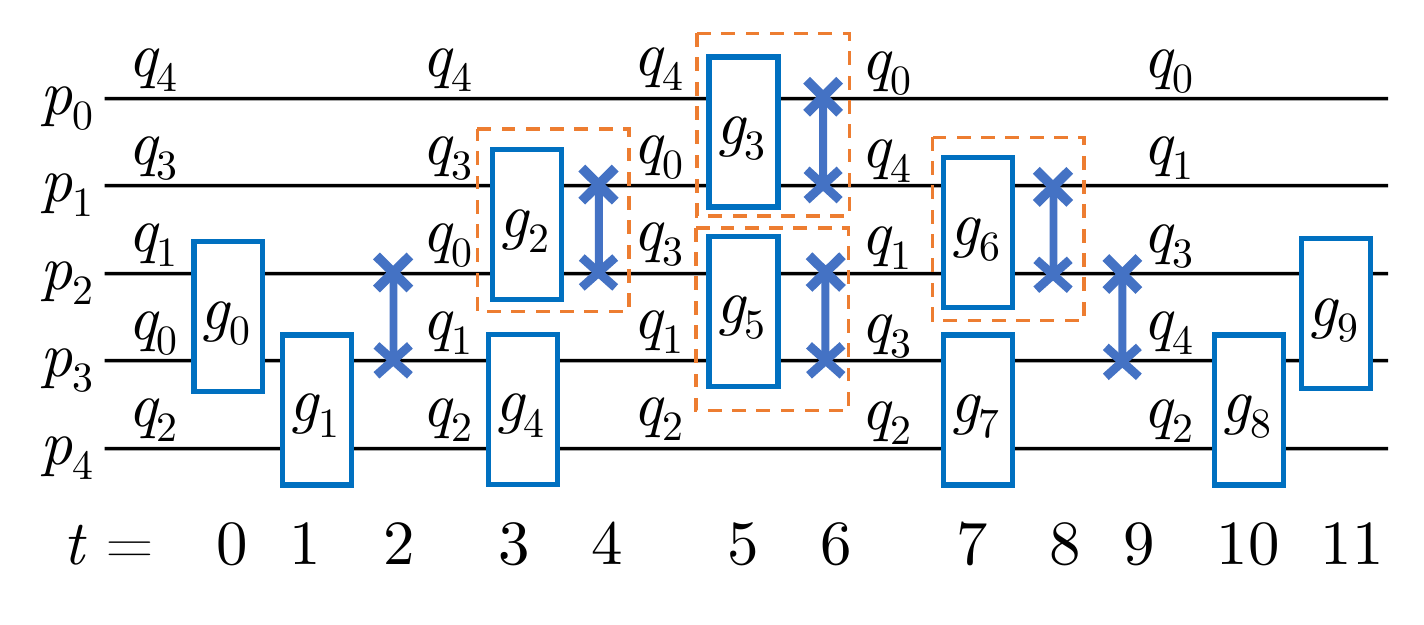}
        \caption{SABRE \cite{asplos19-li-ding-xie-sabre-mapping} solution with 6 SWAPs and depth 12.
        With post-processing, 4 SWAPs can be absorbed, and the depth becomes 9.}
        \label{fig:eg-sabre}
    \end{subfigure}
    \vfill
    \begin{subfigure}[b]{\linewidth}
    \centering
        \includegraphics[scale=0.6]{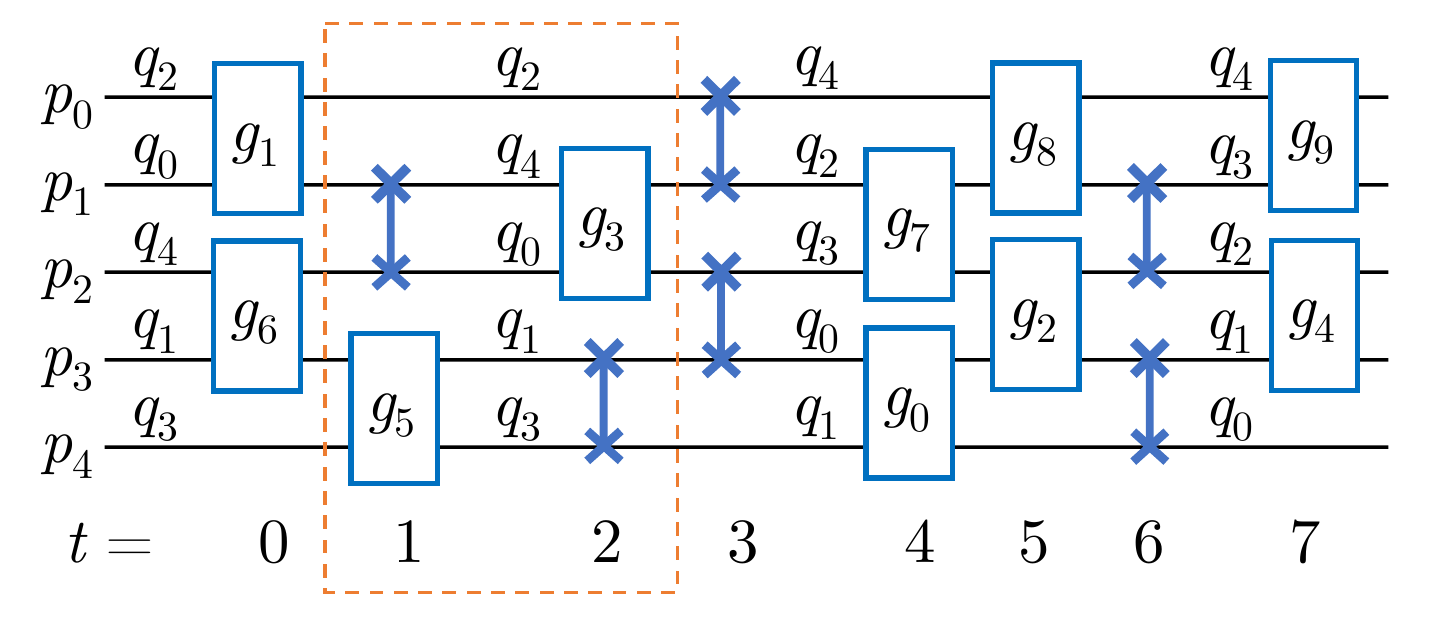}
        \caption{TB-OLSQ \cite{iccad20-tan-cong-optimal-layout-synthesis} solution with 6 SWAPs and depth 8.
        The two steps inside the dashed box can be combined with SWAP absorption as post-processing, then it would have 4 SWAPs and depth 7.}
        \label{fig:eg-olsq}
    \end{subfigure}
    \vfill
    \begin{subfigure}[b]{\linewidth}
        \centering
        \includegraphics[scale=0.6]{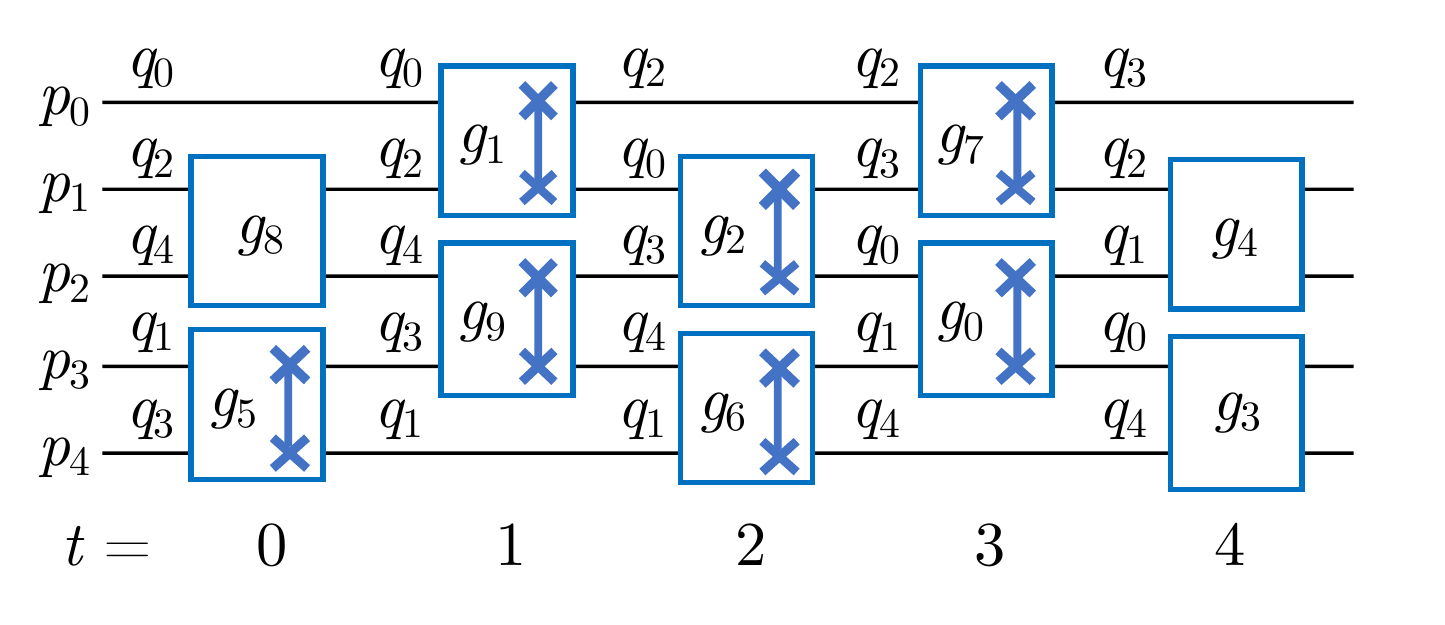}
        \caption{OLSQ-GA solution with no explicit SWAPs and depth 5. The SWAPs inside the boxes are absorbed.}
    \label{fig:eg-optinisq}
    \end{subfigure}
    \caption{Mapping solutions of 5-qubit chemical simulation on a linear architecture.
    Connected crosses are SWAPs. At each time step, which program qubit locates at which physical qubit is shown.}
\end{figure}

When scheduling the gates, there may be some extra degrees of freedom brought by commutation.
The quantum program is really a list of gates.
If two gates act on the same qubit subsequently, their execution order cannot be changed.
We call this a dependency of the latter gate on the former gate, e.g., $g_4$ and $g_0$ subsequently act on $q_1$, so there is a dependency of $g_4$ on $g_1$.
In Fig.~\ref{fig:eg-sabre}, we show a solution by SABRE \cite{asplos19-li-ding-xie-sabre-mapping} which does not exploit the commutations, i.e., all the dependencies are respected.
However, in our simulation example, there are many commutation relations, i.e., we can change the order of gates from the order specified in the program, which means more opportunities for depth and SWAP optimization.

Fig.~\ref{fig:eg-olsq} illustrates a mapping solution by TB-OLSQ \cite{iccad20-tan-cong-optimal-layout-synthesis} with consideration of commutation.
To make the illustration of the mapping easier to read in Fig.~\ref{fig:eg-olsq} (and \ref{fig:eg-optinisq}), we annotate each wire with the logical qubit it refers to, at each step, before the gates scheduled for that particular step.
If the mapping is unchanged from the previous step, then it is omitted.
At time 0, there are two gates $g_1$ and $g_6$.
According to the program in Fig.~\ref{fig:eg-program}, $g_1$ should act on qubit $q_0$ and $q_2$ that are mapped to $p_1$ and $p_0$ at time 0, which agrees with where $g_1$ is.
When advancing to time 1, there is a SWAP gate on $(p_1,p_2)$.
This changes the mapping of $q_0$ to $p_2$ and the mapping of $q_4$ to $p_1$.

\subsection{Metric of Mapping Solution Quality} \label{ssec:analysis-metric}
As mentioned in Sec.~\ref{sec:introduction}, every gate or idleness may introduce error.
The former is captured by gate fidelity and the latter is captured by decoherence factor.
With the assumption of stochastic error, a common fidelity model is just the product of all gate fidelity \cite{nature19-google-quantum-supremacy}.
Since the mapping process only insert SWAPs and does not reduce the original gates, the total fidelity is monotonic to the number of inserted SWAPs.
We mentioned the notion of dependency in quantum programs (or `quantum circuits').
The same notion applies to mapping solutions like Fig.~\ref{fig:eg-olsq}, e.g., $g_{5}$ acts on $p_3$ after $g_{6}$, so $g_5$ depends on $g_6$.
The SWAP at time 2 on edge $(p_3,p_4)$, in turn, depends on $g_5$.
With these chains of dependencies, we can define the depth of the circuit as the length of the longest dependency chain, which is also the minimum of total number of time steps a quantum program can be scheduled.
With the same number of gates, lower depth means less idleness, thus less decoherence.
In summary, we would like mapping solutions with a low number of SWAPs and low depth.

\subsection{``Free Lunch'' for Mapping: SWAP Absorption} \label{ssec:analysis-absorption}
Because of the generality of $U(4)$ gates, we can leverage SWAP absorption to reduce explicit SWAPs and depth.
Suppose a gate $W$ acts on two qubit $p_i$ and $p_j$.
Immediately before or after $W$, a SWAP on $p_i$ and $p_j$ is inserted.
We can actually compute the matrix of $\text{SWAP}\cdot W$ and, after the mapping, decompose the updated matrix.
This way, the updated gate still has the decomposition in Fig.~\ref{fig:kak}, just with different single-qubit gates, which means the SWAP is absorbed into $W$ with practically no cost.
In some literature, this process is called `implementing a mirrored gate' \cite{qst21-ibm-qv64}.

In our simulation example, the solution in Fig.~\ref{fig:eg-olsq} produced by OLSQ \cite{iccad20-tan-cong-optimal-layout-synthesis} is optimal with 6 SWAPs without consideration of absorption.
There is an opportunity to reduce 2 SWAPs in the dashed box, with the absorption of the SWAP before $g_{3}$ and the SWAP after $g_5$.
For $g_5$, we can compute the product
\begin{equation}
    \text{SWAP} \cdot \text{fSim} = \begin{bmatrix}
    1 &0 &0 &0 \\
    0 &-i\sin\theta &\cos\theta &0 \\
    0 &\cos\theta &-i\sin\theta &0 \\
    0 &0 &0 &-e^{-i\phi} \\
    \end{bmatrix},
\end{equation}
and then pass this new matrix to KAK decomposition subroutine.
(Specifically in chemical simulation, the SWAP gate is different from the normal form in Eq.~\ref{fig:two-qubit} in that the bottom right element is $-1$ instead of $1$, but this does not affect the SWAP absorption technique.)
The original fSim gate can be decomposed in the form of Fig.~\ref{fig:kak}, and the new matrix is still in this form, just with different single-qubit gates in Fig.~\ref{fig:kak}.
In this sense, the absorbed SWAP has been performed with \textit{no cost}. 

Fig.~\ref{fig:eg-optinisq} shows a mapping solution by OLSQ-GA, the tool to be presented in this paper, that explores SWAP absorption automatically as part of the mapping process.
It makes use of 6 absorbed SWAPs and no explicit SWAPs.
The achieved depth is 5, which is better than post-processing solution shown in Fig.~\ref{fig:eg-olsq}.

\section{Formulation of OLSQ-GA} \label{sec:formulation}
In this section, we present optimal layout synthesizer for quantum computing with gate absorption, OLSQ-GA, that formulates the mapping problem with SWAP absorption into an SMT optimization problem \cite{tacas08-demoura-bjorner-z3-smt-solver}.
There are two inputs to the program as in Fig.~\ref{fig:eg-problem}: the quantum program consisting of two-qubit gates to map like shown in Fig.~\ref{fig:eg-program}, and the coupling graph of the architecture like shown in Fig.~\ref{fig:eg-arch}.
The objective of OLSQ-GA is to find a solution with optimal depth or SWAP count as expressed in the following subsection.
It is also possible to set the objective to other quantities built from the variables.

\subsection{Variables} \label{ssec:formulation-variables}
There are 4 groups of variables in OLSQ-GA: mapping, spacetime coordinates, absorbed SWAP, and explicit SWAP.
The total number of variables is $|Q|T+2|G|+2|E|T$, where $|Q|$ is the number of qubits in the program, $T$ is the number of time steps, $|G|$ is the number of gates, and $|E|$ is the number of edges in the coupling graph.
We use $q$ to represent program qubits, $p$ for physical qubits, and $e$ for edges in the coupling graph.
We shall use the example in Fig.~\ref{fig:eg-optinisq} for illustration throughout this section.

The mapping variables $\pi_{qt}=p$ means that, at time $t$, program qubit $q$ is mapped to physical qubit $p$, e.g., $\pi_{q_0\ 0}=p_{0}$ and $\pi_{q_1\ 0}=p_{3}$.

The spacetime coordinates of gate $g$ $(t_g,x_g)=(t,e)$ means that $g$ is scheduled at time $t$ and locates on edge $e$ in the coupling graph, e.g., the spacetime coordinates for $g_{0}$ is $(3,e_2)$ where $e_2=(p_2,p_3)$.

A set of absorbed SWAP binary variables $\alpha_{et}$'s are introduced.
If $\alpha_{et}=1$, then there is an absorbed SWAP on edge $e$ at time $t$, e.g., $\alpha_{e_3\ 0}=1$ since there is a SWAP absorbed by $g_{5}$ on edge $e_3=(p_3, p_4)$ at time 0.

Similarly, a set of explicit SWAP binary variables $\sigma_{et}$'s are introduced.
$\sigma_{et}=1$ if and only if there is an explicit SWAP on edge $e$ at time $t$.
There is no explicit SWAP in Fig.~\ref{fig:eg-optinisq}, but in Fig.~\ref{fig:eg-olsq}, $\sigma_{e_1\ 1}=1$ since there is a SWAP on edge $e_1=(p_1,p_2)$ at time 1.

With these variables, the optimization objectives can be easily expressed.
Depth is defined as the largest time coordinate of any gate, $T=\max_g t_g$; SWAP count is the sum of all explicit SWAP variables, $S=\sum_{e,t} \sigma_{et}$; an estimation of fidelity can be the product of a decoherence factor with all gate fidelity
\begin{equation} \label{eq:fidelity}
    f=e^{-\frac{|Q|\cdot T - 2(|G| + S)}{|Q|\cdot T_0}} f_U^{|G|+S},
\end{equation}
where $|Q|$ is the number of program qubits, $T$ is the depth, $|G|$ is the number of gates, $S$ is the SWAP count, and $T_0$ and $f_U$ are hardware factors.
$T_0$ is the decoherence time of a qubit divided by the duration of a $U(4)$ gate, and $f_U$ is the fidelity of a $U(4)$ gate.
In physics, decoherence is characterized by an exponential decay with respect to time.
So, in Eq.~\ref{eq:fidelity}, on the power of the $e$ is the negation of the ratio between the total idle time and the total coherence time.

\subsection{Constraints} \label{ssec:formulation-constraints}
There are five sets of constraints: dependencies, mapping implied by spacetime coordinates, no overlaps, SWAP absorption, and mapping transformation

\textit{Dependencies:} as mentioned in Sec.\ref{ssec:mapping-problem}, e.g., $t_{g_{4}}>t_{g_{0}}$ and $t_{g_{4}}>t_{g_{1}}$.
However, if there is a region in the quantum program where all the gates commute with each other, we can simply change the larger-than relation $>$ to non-equality $\neq$.
Since the simulation gates commute, the actual constraints are $t_{g_{4}}\neq t_{g_{0}}$ and $t_{g_{4}}\neq t_{g_{1}}$.

\textit{Mapping implied by spacetime coordinates:} when gate $g$ acts on program qubit $(q,q')$ at time $t$ on edge $e=(p,p')$,
\begin{equation} \label{eq:mapping-implied-by}
\begin{split}
    &t_g==t \ \wedge\ x_g==e \ \ \Rightarrow \\ &\left(\pi_{qt}==p \ \wedge\ \pi_{q't} == p' \right) \vee \left(\pi_{qt}==p'\ \wedge\ \pi_{q't} == p\right).
\end{split}
\end{equation}
The left-hand side checks the spacetime coordinates of the gate $(t,e)$, while the right-hand side means that, at this time, the mapping of $q$ and $q'$ must be the two physical qubits on $e$, e.g., since $g_{8}$ is at time $0$ and on edge $e_1=(p_1,p_2)$, its two program qubits $q_2$ and $q_4$ should be mapped to $p_1$ and $p_2$.
When we specify an edge with two vertices, there are two possibilities $\pi_{q_2\ 0}=p_1$ and $\pi_{q_4\ 0}=p_2$, or $\pi_{q_2\ 0}=p_2$ and $\pi_{q_4\ 0}=p_1$, which is how the righthand side of Eq.~\ref{eq:mapping-implied-by} got its form.
In our example, the former case is true.

\textit{No overlaps:} there are only two types of gates in our mapping solution, $U(4)$ gates from the program and the explicit SWAPs.
The $U(4)$ gates cannot overlap with each other by the dependency constraints, so we only need to consider the overlaps between $U(4)$ gates and SWAPs, and among SWAPs themselves.
For two incident edges $e$ and $e'$, any gate $g$, and any time $t$, 
\begin{align}
    \sigma_{et}==1 \ \ &\Rightarrow \ \ \sigma_{e't}==0, \\
    t_g==t\ \wedge\ x_g==e \ \ &\Rightarrow \ \ \sigma_{e't}==0,
\end{align}
e.g., there can be no explicit SWAP on edge $e_0=(p_0,p_1)$ or $e_2=(p_2,p_3)$ at time 0 since there is a gate $g_{8}$ on an overlapping edge $e_1=(p_1,p_2)$ at time 0.
In Fig.~\ref{fig:eg-olsq}, there is a SWAP scheduled at time 1 on $e_1$, so there cannot be any other SWAPs or $U(4)$ gates on overlapping edges $e_0$ or $e_2$ at time 1.

\textit{SWAP absorption:} without constraints, an absorbed SWAP can happen on any edge at any time, which is clearly not possible.
If there is an absorbed SWAP on edge $e$ at time $t$, there should also be some $U(4)$ gate, i.e., for any time $t$ and edge $e$,
\begin{equation}
    \alpha_{et}==1\ \ \Rightarrow \ \ \bigvee_g \left( t_g==t\ \wedge\ x_g==e \right),
\end{equation}
e.g., if there is an absorbed SWAP on $e_3=(p_3,p_4)$ at time $0$, then there must be a gate ($g_{5}$ in our example) having spacetime coordinates $(0,e_3)$.

\textit{Mapping transformation:} there are two sources of change for the mapping solution: absorbed and explicit SWAPs.
If either one of them is 1, we deduce the new mapping from the old mapping, i.e., for any qubit $q$, any time $t$, and any edge $e=(p,p')$,
\begin{equation}
    \pi_{qt}==p\ \wedge\ \left(\sigma_{et} == 1\ \vee\ \alpha_{et}==1 \right) \ \ \Rightarrow \ \ \pi_{q\  t+1}=p',
\end{equation}
e.g., $q_1$ is mapped to $p_3$ at time 0, but there is an absorbed SWAP on edge $e_3=(p_3,p_4)$.
As a result, at time 1, $q_1$ is mapped to $p_4$.
Similarly, the mapping of $q_3$ changes from $p_4$ to $p_3$ at time 1.
Thus, $g_{9}$ acting on $q_3$ and $q_4$ can be executed at time 1 on $e_2=(p_2,p_3)$, but not at time 0 because of the mapping.
On the other hand, if there are no SWAPs, absorbed or explicit, on any edge going into the current physical qubit, the mapping remains unchanged from $t$ to $t+1$, i.e.,
\begin{equation}
    \pi_{qt}\ \wedge\left(\sum_{p\in e} \sigma_{et} == 0\right) \wedge \left( \sum_{p\in e} \alpha_{et} == 0 \right) \ \ \Rightarrow \pi_{q\ t+1}=p,
\end{equation}
e.g., at time 1, there is neither absorbed nor explicit SWAP on $e_0$ or $e_1$, so the mapping of $q_2$ remains at $p_1$.

\section{Application: Mapping QAOA for 3-Regular Graphs} \label{sec:qaoa}

\begin{figure}[bht]
    \centering
    \includegraphics[scale=0.5]{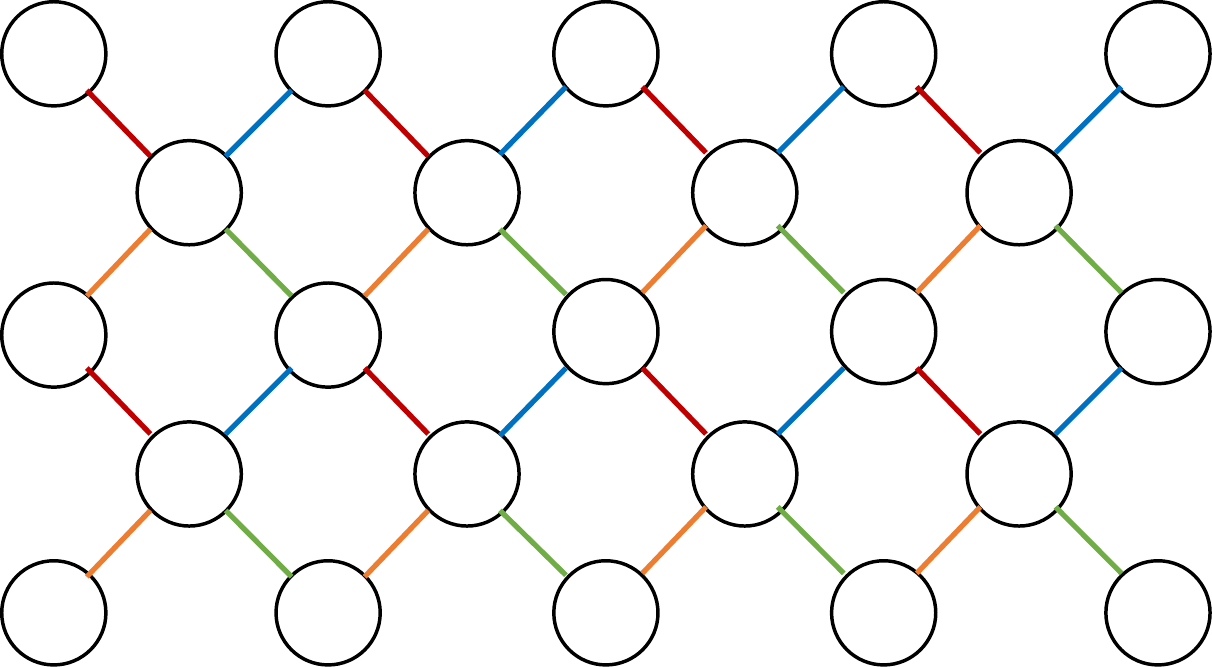}
    \caption{Part of Google Sycamore \cite{natphys21-google-qaoa}. Four different colors represent four maximal matchings of the coupling graph.}
    \label{fig:sycamore}
\end{figure}

QAOA can be adapted to many optimization problems.
One of the promising candidates is the MAXCUT problem on 3-regular graphs \cite{algorithms19-hadfield-wang-ogorman-rieffel-venturelli-biswas-qaoa}.
A QAOA program typically consists of $p$ iterations ($p\in\mathbb{N}$), and each iteration consists of two stages: \textit{phase-splitting} and \textit{mixing}.
The mixing stage only has single-qubit gates, so there is no mapping problem.
The phase-splitting stage, however, presents an interesting mapping problem.
Specifically, for the MAXCUT problem on a graph $G=(V,E)$, each qubit encodes a vertex, and we need to apply a two-qubit gate on every edge of $G$.
These gates are all commutable.
A state-of-the-art experimental work used a heuristic compiler \cite{qst20-sivarajah-dikes-cowtan-simmons-edgington-duncan-tket-compiler-nisq} and coupling graph in Fig.~\ref{fig:sycamore}, but the quality of the result dropped quickly with increasing problem sizes.

\begin{figure}[bht]
    \centering
    \includegraphics[scale=0.6]{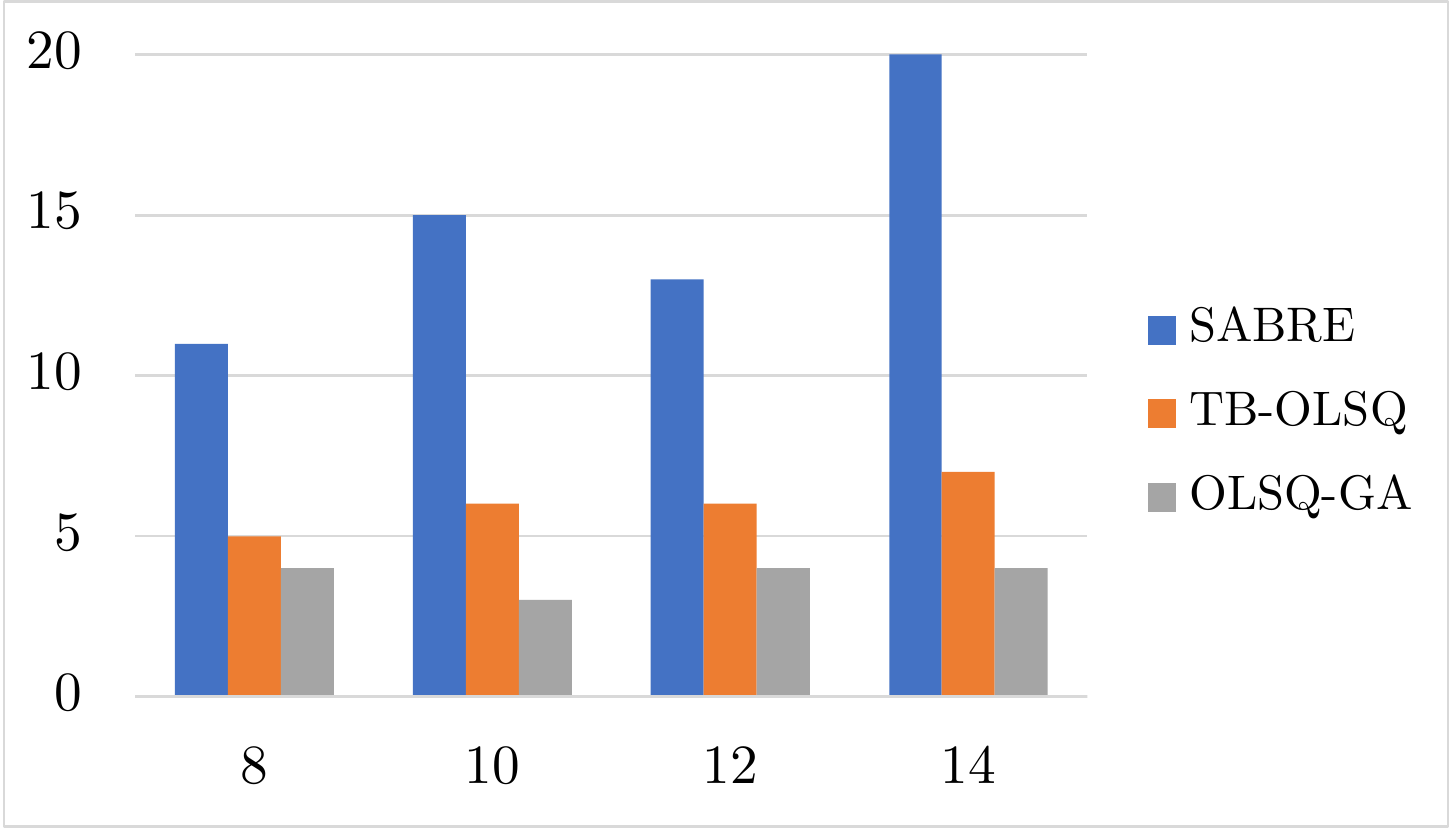}
    \caption{Depth by three mapping approaches.}
    \label{fig:qaoa-depth}
\end{figure}

\begin{figure}[bht]
    \centering
    \includegraphics[scale=0.6]{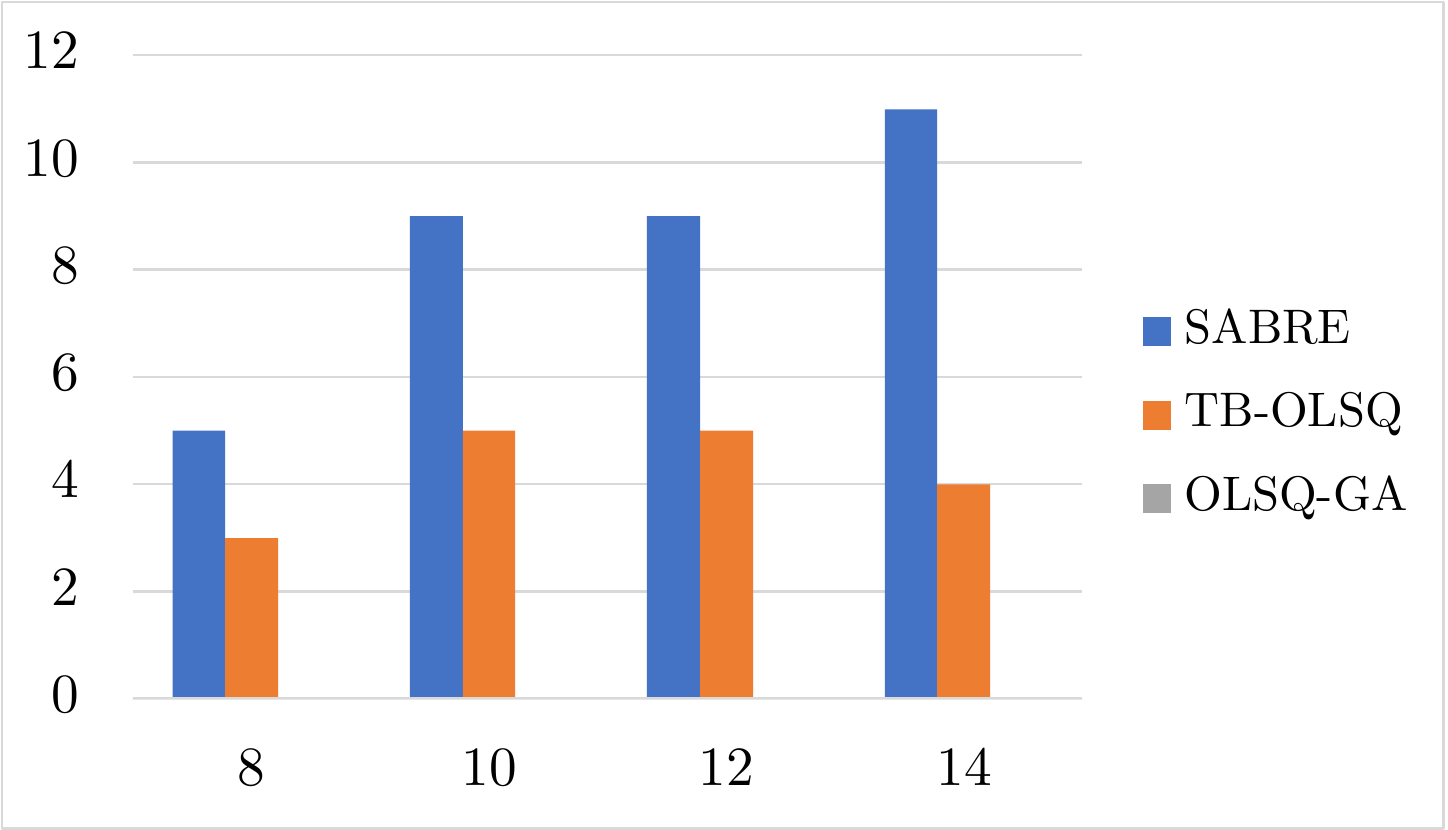}
    \caption{SWAP count by three mapping approaches. Note that OLSQ-GA managed to insert no explicit SWAP gates, so there are no gray bars in the graph above.}
    \label{fig:qaoa-swap}
\end{figure}

\begin{figure}[bth]
    \centering
    \includegraphics[scale=0.6]{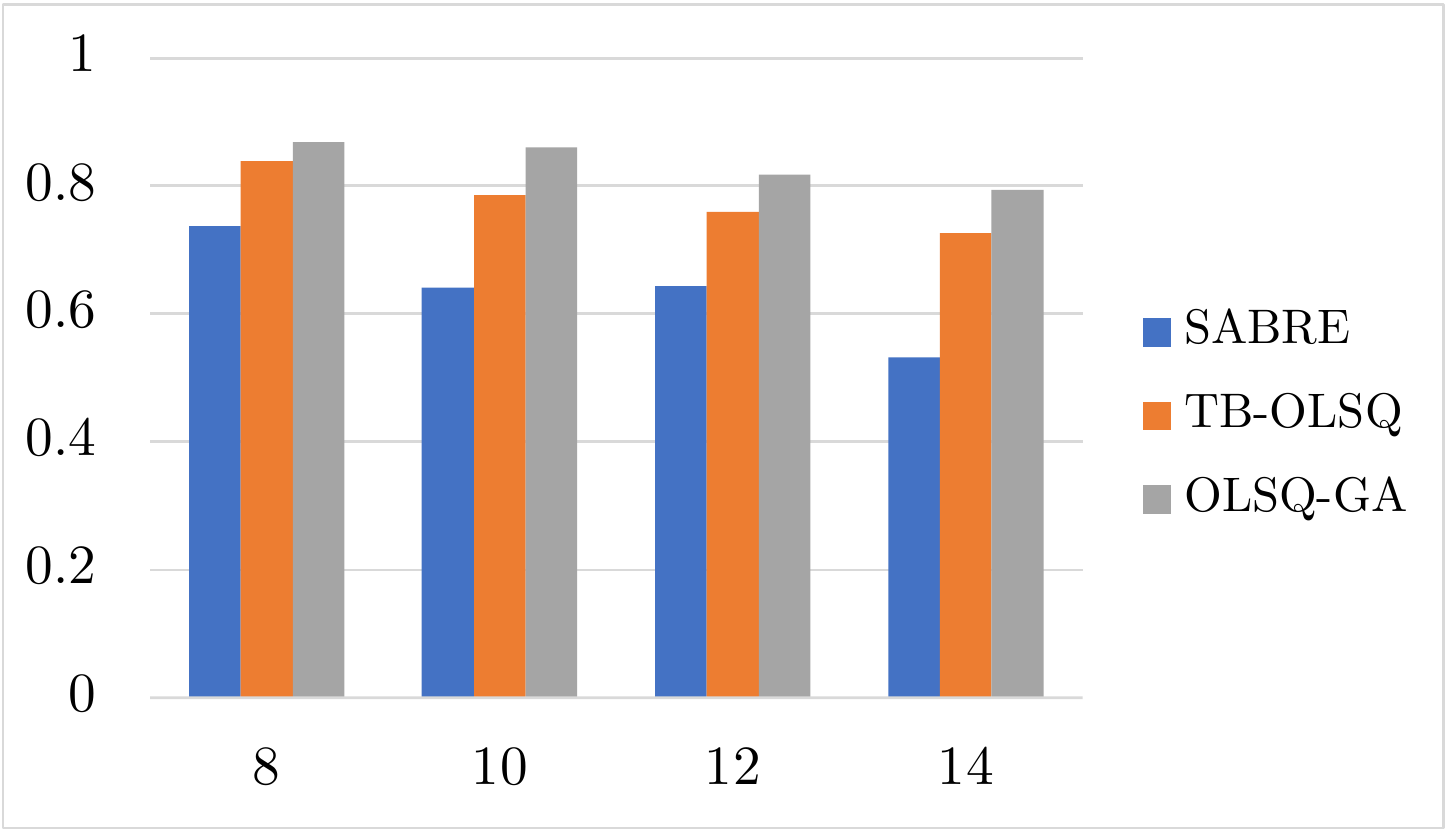}
    \caption{Fidelity by three mapping approaches.}
    \label{fig:qaoa-fidelity}
\end{figure}

We implemented OLSQ-GA\footnote{\url{https://github.com/UCLA-VAST/OLSQ/tree/GateAbsorption}} with Z3 SMT solver \cite{tacas08-demoura-bjorner-z3-smt-solver} and generated four 3-regular graphs of sizes 8, 10, 12, and 14 with the NetworkX package \cite{lanl08-hagberg-swart-chult-networkx} as the benchmark, similar to the setting in Google's experimental work \cite{natphys21-google-qaoa}.
We evaluated OLSQ-GA against two tools with the same benchmark: SABRE \cite{asplos19-li-ding-xie-sabre-mapping}, and TB-OLSQ \cite{iccad20-tan-cong-optimal-layout-synthesis}.
Although SABRE is not exactly what was used in \cite{natphys21-google-qaoa}, it is also considered to be state-of-the-art for heuristic mapping \cite{qst21-ibm-qv64}.
TB-OLSQ uses an optimal approach but does not take the gate absorption into consideration.
We set the number of SWAPs as the objective in OLSQ-GA.
The depth, SWAP count, and fidelity of the mapping solutions for the phase-splitting stage of a single iteration in QAOA are shown in Fig.~\ref{fig:qaoa-depth}, Fig.~\ref{fig:qaoa-swap}, and Fig.~\ref{fig:qaoa-fidelity} respectively.
The fidelity is estimated by Eq.~\ref{eq:fidelity} with slightly optimistic parameters $T_0=50$ and $f_U=0.99$ \cite{misc-ibm-quantum-experience}, which means that decoherence time is 50X the $U(4)$ gate duration, and each $U(4)$ gate fidelity is $99\%$.
As we can see, the heuristic tool, without consideration of the SWAP absorption or commutation, returns solutions with the highest depth and SWAP counts, and thus the lowest fidelity.
The results of TB-OLSQ are already significantly better than the heuristic results.
OLSQ-GA performs the best of all three.
Compared to TB-OLSQ, it reduces depth by up to 50.0$\%$ and SWAP count by 100$\%$ while improving fidelity by 9.45$\%$.
Compared to SABRE, it reduces depth by up to 80.0$\%$, SWAPs by 100$\%$ and improves fidelity by up to 49.1$\%$.

Note that all the mapping solutions for one iteration can easily extend to multiple iterations: we can simply reverse the order of all the gates and append this reversed circuit as the second iteration.
Of course, in the new iteration, there are different parameters in the gates, but the mapping problem can be solved just for one iteration.
This way, the final mapping of all the odd iterations is the same as the final mapping of the first iteration, and the final mapping of the even iterations is just the initial mapping.
The total fidelity of all iterations is the product of fidelity of each, so the total fidelity would be exponential to the single-iteration fidelity.
Since the QAOA circuits with more iterations contain the QAOA circuits with less iterations, in the ideal case, the quality of QAOA results should increase as the number of iteration $p$ increases.
However, in the leading experimental work \cite{natphys21-google-qaoa}, such a trend is only observed on hardware-efficient graphs, not the generated 3-regular graphs like what we use in this paper.
Without quantum error correction, the fidelity of the whole circuit is only going to decrease as the number of gates increases.
However, the quality of the QAOA results is not proportional to the circuit fidelity, which is why some improvements are still observed.

\begin{figure}[bth]
    \centering
    \includegraphics[scale=0.6]{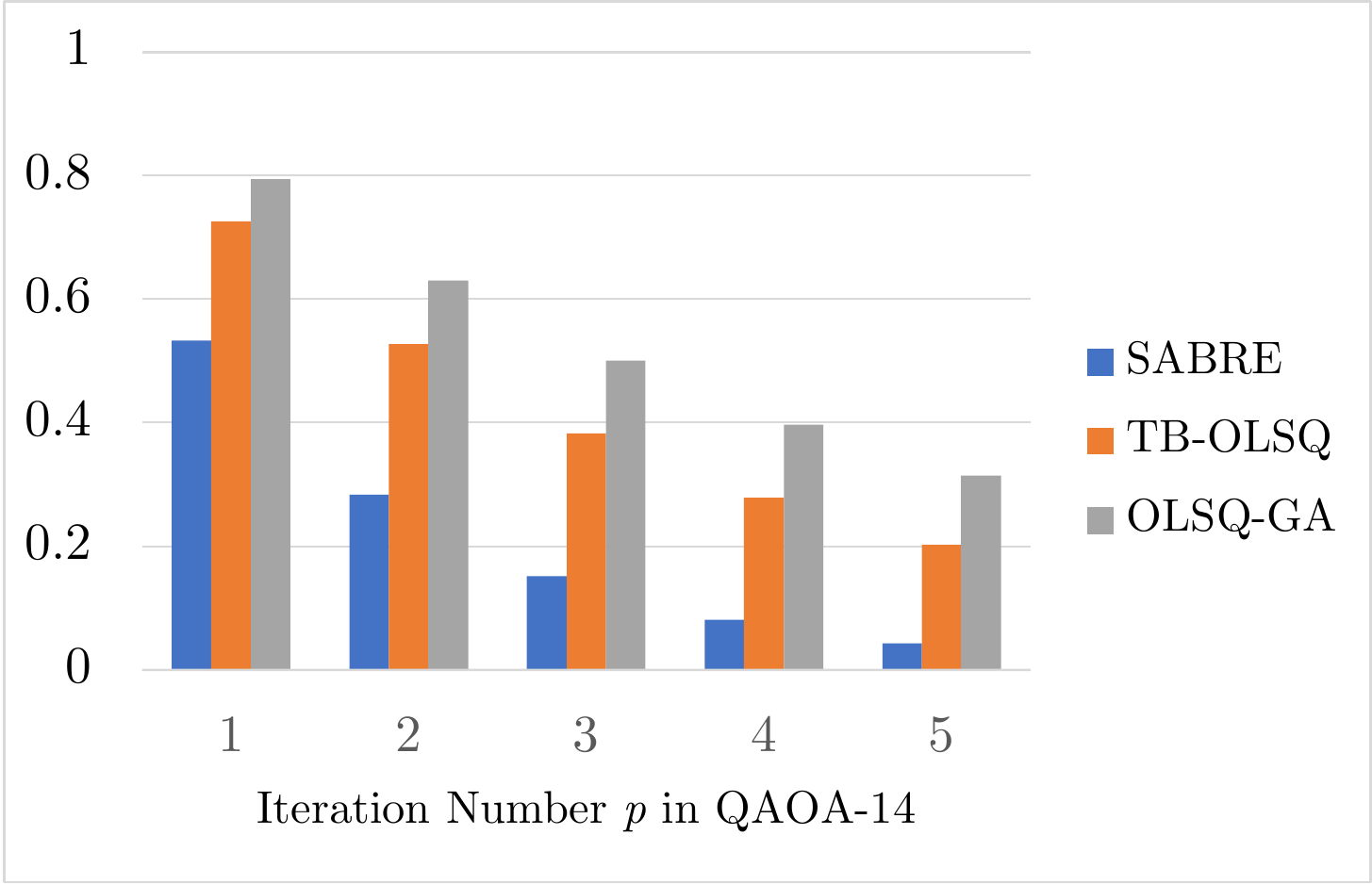}
    \caption{Fidelity of multiple iterations of QAOA-14 (QAOA for the 3-regular graph with 14 vertices) by three mapping approaches.}
    \label{fig:qaoa-fidelity-iter}
\end{figure}

Apart from still-low gate fidelity, we believe one of the reasons is the sub-optimal compilation for mapping, e.g., authors of \cite{natphys21-google-qaoa} report that the depth of the heuristic mapping solutions for a single QAOA iteration is approximately the size of the 3-regular graph.
In comparison, the depth of OLSQ-GA results stays as a constant (3 or 4), which is way less than the size of the graphs (8 to 14), and the same or lower than the hardware-efficient graphs.
This suggests that, using OLSQ-GA, the existing hardware capability could also demonstrate improvements with more iterations for 3-regular graphs.
Fig.~\ref{fig:qaoa-fidelity-iter} shows the fidelity of three mapping approaches with up to 5 iterations. As the number of iterations increases, the advantage of OLSQ-GA becomes more visible: compared to TB-OLSQ, it gains fidelity by 30.5$\%$ for 3 iterations, and 55.9$\%$ for 5 iterations; compared to SABRE, the improvements are 231$\%$ are 636$\%$.

\section{Solution Space Reduction} \label{sec:analysis}
The solution space of SMT optimization for OLSQ-GA may be reduced with adding more constraints, thus speeding up the solver.
We shall present two techniques in this section: using alternating matchings pattern and setting initial mapping.

\subsection{Analysis on Optimal Mapping Solutions} \label{ssec:analysis-matching}
In graph theory, matching is a set of pair-wise nonadjacent edges, none of which are self-loops.
The $U(4)$ gates executing at the same time step $t$ consist of a matching on the coupling graph, $M_t$.
In general, we have the following theorem.
\begin{theorem} \label{theorem1}
In a depth-optimal mapping solution, for any $t$, $M_t \cup M_{t+1}$ cannot be a matching for the coupling graph.
\end{theorem}
\textit{Proof:}
If $M_t \cup M_{t+1}$ is a matching in a depth-optimal mapping solution $S$, we can move all of $M_{t+1} \setminus M_t$ to time $t$, and absorb all of $M_{t+1} \cap M_t$ to the corresponding gates in $M_t$.
(Note that, due to the generality of $U(4)$, not only SWAPs, but also any $U(4)$ gate can be absorbed into another $U(4)$ gate.) 
Then we get a new mapping solution $S'$ where the time step $t+1$ is not needed, which contradicts to the fact that $S$ is depth-optimal. $\hfill\square$

For example, in Fig.~\ref{fig:eg-olsq}, $M_1=M_2=\{(p_1,p_2), (p_3,p_4\}$, so $M_1\cup M_2=M_1$ is a matching of the coupling graph, Fig.~\ref{fig:eg-arch}.
The gates at time 2 can be absorbed into gates at time 1.

Analyzing the solution in Fig.~\ref{fig:eg-optinisq}, we can observe a pattern: the gates alternate between two matchings $M_0=\{(p_1,p_2),(p_3,p_4)\}$ and $M_1=\{(p_0,p_1),(p_2,p_3)\}$.
In fact, for linear architecture, we can formalize this observation.
We call the edges in $M_0$ even edges, and edges in $M_1$ odd edges.
\begin{corollary} \label{corollary1}
For the mapping problem of programs with commutation to a linear architecture with coupling graph $G=(P, E)$ where $P$ is the set of physical qubits, and $E=\{(p_{i},p_{i+1})| i=0,..., |P|-2\}$, there is always an depth-optimal mapping solution such that the time steps alternate between sets of even edges and sets of odd edges.
\end{corollary}
\textit{Proof:}
From Theorem~\ref{theorem1}, $M_t \cup M_{t-1}$ cannot be a matching.
For the linear architecture, this means that there are both odd and even edges in $M_t \cup M_{t-1}$, since with only odd or even edges, $M_t \cup M_{t-1}$ would still be a matching.
By absorbing and moving, we can always build new time steps $t-1$ and $t$ such that $t-1$ only has gates on even edges, and $t$ only has gate on odd edges.
Since $M_t \cup M_{t-1}$ has both even and odd edges, none of the two new time steps can be empty.
As a result, we have constructed a new optimal solution satisfying the alternating matchings pattern with the same depth. $\hfill\square$

\subsection{Implementing Alternating Matchings Pattern} \label{ssec:formulation-matching}

\begin{table}[bt]
    \centering
    \begin{tabular}{|ll|c|ccc|}
    \hline
    Problem & Obj. &Baseline & Match. & Init. &Both \\
    \hline
    5-qubit simulation & SWAP &4.74E0 &1.40X &2.58X  &2.44X\\
    \hline
    QV64 &Depth &2.40E2 &6.35X &5.00X  &8.86X\\
    &SWAP &8.50E3 &95.4X &53.0X  &272X\\
    \hline
    QAOA-14, Sycamore &Depth &1.65E5 &8.41X* &522X* &* \\
    \hline
    \end{tabular}
    \caption{OLSQ-GA speedup with extra constraints. The architectures for simulation and QV64 are linear. The architecture for QAOA is shown in Fig.~\ref{fig:sycamore}. Baseline is the runtime in seconds. `Obj.' means objective. `Init.' means fixing initial mapping. `Match.' means using alternating matchings pattern. The asterisk (*) means that the mapping solution using the corresponding technique(s) may not be optimal. However, the depth of the two cases with data shown above matched the depth certificate as in Sec.~\ref{ssec:certificate}, so these solutions are indeed optimal.}
    \label{tab:speedup}
\end{table}

For linear architecture, Corollary~\ref{corollary1} leads to a great reduction in solution space of the mapping problem without loss of optimality.
In OLSQ-GA formulation, this can be achieved by assigning values to many explicit SWAP and space variables for $U(4)$ gates.
For all $(t,e_k)$ such that $(t-k)\mod 2 == 1$, and all gate $g$,
\begin{align} \label{eq:matchings}
    &\sigma_{e_k\ t}=0,\\
    &t_g == t\ \ \Rightarrow\ \  x_g \neq e_k.
\end{align}
Note that a single $=$ means assigning value to the variable.
These constraints make sure that there are only gates on even edges at time 0, 2, 4, ... 
And there are only gates on odd edges at time 1, 3, 5, ...
If there is an even number of edges in the linear architecture, these constraints suffice.
However, if there is an odd number of edges, we may need to try another case with $(t-k)\mod 2 == 0$ instead of 1.
The two matchings have a different number of edges, so it matters which one we start from.
Taking the better result of the two cases, we derive the optimal result.

Since we have fixed some variables and added more constraints, the solution space for the solver to explore is smaller, which results in a faster runtime.
Some speedup results are shown in Table~\ref{tab:speedup}.
When mapping the QV64 circuit \cite{qst21-ibm-qv64}, alternating matchings bring 95.4X speedup.

For generic architectures, it is more complex.
For example, for a 2D architecture like Fig.~\ref{fig:sycamore}, there are four maximal matchings, shown in different colors, that are mutually disjoint in a sense similar to $M_0$ and $M_1$.
Ref.~\cite{natphys21-google-qaoa} alternates among these matchings to schedule a single QAOA iteration for hardware-efficient graphs: at each time step, a group of gates with the same color are executed.
However, the optimal mapping solution for other quantum circuits may use other possible ordering of these four matchings, or even other possible matchings.

\subsection{Depth Certificate} \label{ssec:certificate}

There is a generic case where we can guarantee optimal depth even with heuristics: we can run two OLSQ-GA instances with the heuristics turned on and off.
The two instances start with a certain maximal depth.
If the current maximal depth is too low to yield any solution, OLSQ-GA would increase the maximal depth and start over.
The exact instance explores a larger solution space, so its runtime is longer.
Meanwhile, it can output what is the maximal depth currently being explored, e.g., 4, which serves as a certificate that no solutions with depth less than 4 can be found.
Now, if the heuristic instance returns a mapping solution with depth 4, which takes less time than the exact instance, then the solution is optimal because of the depth certificate by the exact instance.
In Table~\ref{tab:speedup}, we also report the speedup of the 14-qubit QAOA with alternating matchings, using depth as the objective.
For SWAP count, the optimality argument would be harder to guarantee.
However, if the heuristic solution does not contain any explicit SWAPs, then it is optimal with respect to SWAP count.

\subsection{Setting Initial Mapping}
Another technique to reduce solution space is to set initial mapping.
If there are not too many qubits, we can send instances with different initial mappings to different cores and perform the solving in parallel.
For problems with a strong symmetry, we can set initial mapping to ``break'' some symmetry without loss of optimality, e.g., in the 5-qubit all-to-all chemical simulation, we can use arbitrary initial mapping, and the speedup is 2.58X.
As implementation, we can add these constraints to OLSQ-GA:
\begin{equation} \label{eq:mappings}
    \pi_{q_i\ 0} = p_i \ \ \text{for }i=0,...,4
\end{equation}

If the gates are not commutable like in QV64, the initial mapping should enable some gates to execute since the SWAPs before all the $U(4)$ gates can simply be left out and we set the initial mapping to be directly whatever mapping it is after these ``prelude'' SWAPs.
QV also has a special property that its first time step is a maximal matching consisting of $\lfloor n/2 \rfloor$ gates. Being exhaustive, we can let each core in a computational cluster try one of the $ \lfloor n/2\rfloor !\ 2^n/2$ possible initial mappings.
The factorial term is the number of mappings from the gates to edges on the coupling graph.
The exponential term is for both directions of each edge.
At last, note that, if the architecture is 1D and we reflect a mapping solution with respect to the center, we get another solution with the same depth and SWAP count.
Thus, we can divide the possibilities of initial mapping by 2 in Eq.~\ref{eq:mappings}.
Using Sterling's approximation, the asymptotic of this value is $\sqrt{\pi n}(n/e)^n/2$, which is approximately 35$\%$ of all the possible initial mappings $n!$.
For $n=6$, the required core count is 192, which is not too much in distributed computing.
The best solution of all these cases is still guaranteed to be optimal.
We chose one possibility and achieved 53X speedup, as shown in Table~\ref{tab:speedup}.
With both alternating matching and initial mapping, we achieve up to 272X speedup.

We can also use the initial mapping results as the heuristic in Sec.~\ref{ssec:certificate}.
For example, we used TB-OLSQ to derive an initial mapping for the 14-qubit QAOA and use it in OLSQ-GA.
The combined runtime of TB-OLSQ and OLSQ-GA is still 522X faster than the baseline.
However, note that combining alternating matchings and initial mapping may cause issues.
The initial mapper may not produce an alternating matchings solution.
So, it cannot be combined with alternating matchings to produce a depth-optimal solution.

\section{Related Works} \label{sec:related}
There have been multiple studies on reducing depth or SWAP count for qubit mapping.
Two state-of-the-art heuristic search methods in academia,  Ref.~\cite{date18-zulehner-paler-wille-efficient-mapping-ibmqx} and SABRE \cite{asplos19-li-ding-xie-sabre-mapping},  have been incorporated into Qiskit, the software platform of IBM Quantum.
Other noticeable tools from the industry include QUILC \cite{qst20-smith-peterson-ski;beck-davis-rigetti-quilc-compiler} and t$|$ket$\rangle$ \cite{qst20-sivarajah-dikes-cowtan-simmons-edgington-duncan-tket-compiler-nisq}.
Since the size of the NISQ application we can run is still moderate due to the capability of existing hardware, exact and optimal methods are still very valuable to make the most of the hardware.
Ref.~\cite{dac19-wille-burgholzer-zulehner-mapping-minimal-swaph} optimizes the SWAP count using SMT \cite{tacas08-demoura-bjorner-z3-smt-solver}.
Ref.~\cite{asplos21-zhang-hayes-qiu-jin-chen-zhang-time-optimal-mapping} minimizes depth using A* search with an admissible heuristic to guarantee optimality \cite{book09-russell-norvig-artificial-intelligence}.
OLSQ \cite{iccad20-tan-cong-optimal-layout-synthesis} can optimize either depth, SWAP count, or fidelity using SMT.
In a more theoretical flavor, the mapping problem is also termed as `routing via matching' \cite{tqc19-childs-schoute-unsal-circuit-transformation}.
Ref.~\cite{arxiv1905-ogorman-huggins-rieffel-whaley-swap-network-nisq} gives lower-bounding techniques and constructions for some NISQ applications on linear architectures.
In the event of non-uniform gate fidelity \cite{asplos19-tannu-qureshi-variability-aware-policy} or even correlated error \cite{asplos20-murali-mckay-martonosi-javadi-abhari-mitigation-crosstalk}, it is important to map the quantum program to high-fidelity qubits in the architecture.
TriQ \cite{isca19-murali-linke-martonisi-abhari-nguyen-alderete-triq-architecture-studies}, MUQUT \cite{iccad19-bhattacharjee-saki-alam-chattopadhyay-ghosh-muqut-mapping}, Ref.~\cite{asplos20-murali-mckay-martonosi-javadi-abhari-mitigation-crosstalk}, and OLSQ \cite{iccad20-tan-cong-optimal-layout-synthesis} formulate this problem differently and both solve it using the z3 SMT solver \cite{tacas08-demoura-bjorner-z3-smt-solver}.

However, there have been very limited works taking advantage of SWAP absorption and other NISQ properties presented in Sec.~\ref{sec:analysis}.
Ref.~\cite{aspdac19-zulehner-wille-su(4)-compiling} extended the A* search with SWAP absorption by making the cost of a SWAP to be 0 if it is immediately after a $U(4)$ gate.
The most relevant work is a mapping tool from IBM \cite{arxiv2106-nannicini-bishop-gunluk-jurcevic-optimal-mapping-bip, qst21-ibm-qv64}, which formulates the problem in binary integer programming (BIP) and solves it using a proprietary solver, CPLEX.
For the example in \cite{qst21-ibm-qv64}, OLSQ-GA finds an optimal solution with the same quality (depth 11, 8 SWAPs).
On this very example, the solution of Ref.~\cite{aspdac19-zulehner-wille-su(4)-compiling} has depth 15 and 11 SWAPs.
(This is the best case out of 10 trials since its A* algorithm has some randomness.)
In comparison, the solution of SABRE has depth 12 and 9 SWAPs \cite{qst21-ibm-qv64}.

For QAOA and chemical simulation, there is a known optimal mapping solution to the instances with ``all-to-all'' interactions, like Fig.~\ref{fig:eg-problem}.
In this solution, the gates are arranged in alternating matchings, and each with an absorbed SWAP gate.
This corresponds to the most general kind of chemical simulation program, where every qubit has a gate with every other qubit, so the optimal solution has depth $n-1$ with a total of ${n \choose 2}$ two-qubit gates.
Ref.~\cite{prl18-kivlichan-mcclean-wiebe-gidney-aspuru-guzik-chan-babbush-vqe-linear-ansatz} provides more details on this optimal mapping solution.
Such solution also works for QAOA for complete graphs, i.e., the Sherrington-Kirkpatrick model \cite{natphys21-google-qaoa}.
However, for problems with fewer gates than the all-to-all interactions, we may not need ${n \choose 2}$ gates.
In this case, the depth-optimal solution is less structured and OLSQ-GA is helpful to find it.

\section{Conclusion} \label{sec:conclusion}
By analyzing the properties of NISQ applications, we present three techniques improving qubit mapping quality and efficiency: SWAP absorption, commutation, and alternating matchings.
Applying these techniques, we present OLSQ-GA, a mapper that formulates the generalized NISQ mapping problem (with SWAP absorption) using SMT optimization and solves it optimally.
Comparing to state-of-the-art optimal method, we reduce depth by up to 50.0$\%$, SWAP count by 100$\%$.
We improve fidelity by 9.45$\%$ for a single iteration, and 55.9$\%$ for 5 iterations on a set of QAOA instances.
For future directions, it is valuable to 1) devise alternating matchings for different architectures with low loss of optimality, 2) parallelize the solving process by partitioning of solution space or other methods, and 3) assisted by OLSQ-GA, find optimal mapping solutions of important chemical models to simulate on realistic quantum architectures.

\section*{Acknowledgments}
This work is partially supported by NEC under the Center for Domain-Specific Computing Industrial Partnership Program.
We would like to thank Ryan Babbush, Austin Minnich, Garnet Chan, Shi-Ning Sun, and Ruslan Tazhigulov for discussions on quantum chemistry; Petar Jurcevic for discussions on quantum volume circuits; Matthew Harrigan for discussions on QAOA.

\bibliographystyle{IEEEtran}
\bibliography{IEEEabrv,daniel-bochen-tan-certified}

\end{document}